%% file: main.tex
\documentclass[11pt]{article}
\usepackage{fullpage}
\usepackage{amsmath, amssymb, amsfonts, amsthm,mathtools,bbm}
\usepackage{xcolor}
\usepackage{thmtools, thm-restate}

\definecolor{DarkRed}{rgb}{0.1,0.1,0.9}
\definecolor{DarkBlue}{rgb}{0.1,0.1,0.5}

\usepackage{nameref}
\definecolor{ForestGreen}{rgb}{0.1333,0.5451,0.1333}
\definecolor{Red}{rgb}{0.9,0,0}
\usepackage[linktocpage=true,
	pagebackref=true,colorlinks,
		urlcolor=black,
	linkcolor=DarkRed,citecolor=ForestGreen,
	bookmarks,bookmarksopen,bookmarksnumbered]
	{hyperref}
\usepackage[noabbrev,nameinlink]{cleveref}
\crefname{property}{property}{Property}
\creflabelformat{property}{(#1)#2#3}
\crefname{equation}{eq}{Eq}
\creflabelformat{equation}{(#1)#2#3}
\usepackage{thmtools}
\usepackage{bbm}
\usepackage[british]{babel}
\usepackage[autostyle]{csquotes}
\usepackage{multicol}

\usepackage{cite}
\input{macros}

\newtheorem{theorem}{Theorem}[section]

\newtheorem{lemma}[theorem]{Lemma}

\newtheorem{definition}[theorem]{Definition}
\numberwithin{equation}{section}

\renewcommand{\qed}{\nobreak \ifvmode \relax \else
      \ifdim\lastskip<1.5em \hskip-\lastskip
      \hskip1.5em plus0em minus0.5em \fi \nobreak
      \vrule height0.75em width0.5em depth0.25em\fi}

\begin{document}

\title{On Minimizing Generalized Makespan on Unrelated Machines}
\date{}
\author{ 
Nikhil Ayyadevara\thanks{University of Michigan, CS Dept., \url{vsnikhil@umich.edu}} \and 
Nikhil Bansal\thanks{University of Michigan, CS Dept., \url{bansal@gmail.com }. Supported in part by the NWO VICI grant 639.023.812.} \and  Milind Prabhu \thanks{University of Michigan, CS Dept., \url{milindpr@umich.edu}}  }

\maketitle

\input{abstract}

\input{intro}

\input{prelims}
\input{integrality-gap}

\input{reduction}

\input{conclusion}

\bibliographystyle{alpha}
\bibliography{general}
\end{document}

%% file: macros.tex
\newcommand{\m}{m}
\newcommand{\bt}{\beta}
\newcommand{\mach}[2]{#2_#1}
\newcommand{\schedule}{\rho}
\newcommand{\Load}{\mathsf{load}}
\newcommand{\poly}{\text{poly}}
\newcommand{\instance}{\mathcal{I}}
\newcommand{\labels}{L}
\newcommand{\Set}[3]{S_{#1,#2,#3}}
\newcommand{\term}[1]{\psi^{(#1)}}
\newcommand{\lsize}{N}
\newcommand{\razconstant}{c}
\newcommand{\satvars}{t}
\newcommand{\normvar}{\mathbf{v}}

\newcommand{\Top}[1]{\mathsf{Top}_{#1}}

\newcommand{\R}{\mathbb{R}}

%% file: abstract.tex
\begin{abstract}
    
We consider the {\em Generalized Makespan Problem} (GMP)
on unrelated machines, where we are given $n$ jobs and $m$ machines and each job $j$ has
arbitrary processing time $p_{ij}$ on machine $i$. Additionally, there is a general symmetric monotone norm $\psi_i$ for each machine $i$, 
that determines the load on machine $i$ as a function of the sizes of jobs assigned to
it. The goal is to assign the jobs to minimize the maximum machine load.

Recently, Deng, Li, and Rabani \cite{deng2023generalized} gave a $3$ approximation for GMP when the $\psi_i$ are top-$k$ norms, and they ask the question whether an $O(1)$ approximation exists for general norms $\psi$?
We answer this negatively and show that, under natural complexity assumptions, there is some fixed constant $\delta>0$, such that GMP is $\Omega(\log^{\delta} n)$ hard to approximate. We also give an $\Omega(\log^{1/2} n)$ integrality gap for the natural configuration LP.
\end{abstract}

%% file: intro.tex
\section{Introduction}
We consider a question by Deng, Li and Rabani \cite{deng2023generalized} about scheduling jobs to minimize makespan on unrelated machines in the setting of more general norms.
Recall that in the unrelated machines setting, we are given $n$ jobs and $m$ machines where each job $j \in [n]$ has some arbitrary processing time $p_{ij}$ 
on machine $i \in [m]$. Given an assignment of jobs to machines, the load on a machine is the sum of the processing times of all the jobs assigned to it. 
In a seminal result, Lenstra, Shmoys and Tardos \cite{lenstra1990approximation} gave a $2$-approximation for minimizing the makespan (the maximum machine load).
These results were later extended to the problem of minimizing the $\ell_p$-norm of the machine loads \cite{alon1998approximation,azar2005convex,kumar2009unified,makarychev2018optimization}.

In a breakthrough work \cite{chakrabarty2019approximation}, Chakrabarty and Swamy introduced a substantial generalization of the problem to general symmetric monotone norms. Recall that a function $\phi:\R^m\rightarrow \R_{\geq 0}$ is a norm if it satisfies (i) $\phi(u)=0$ iff $u=0^n$, (ii)
$\phi(\alpha u) = |\alpha|u$ for all $u$ and $\alpha \in R$ and, 
(iii) $\phi(u+v) \leq \phi(u) + \phi(v)$. 
We say that $\phi$ is {\em symmetric} if $\phi(u)=\phi(u')$ where $u'$ is some permutation of $u$, 
and it is monotone if $\psi(u) \leq \psi(v)$ for all  $u,v$ satisfying 
$ 0\leq u(i) \leq v(i)$ for all $i\in [m]$.

Surprisingly, they showed that for any arbitrary symmetric monotone norm $\phi:\R^m\rightarrow \R_{\geq 0}$,
that determines the overall objective as a function of the individual machine loads, there is an $O(1)$ approximation. The approximation ratio was subsequently improved to $4+\epsilon$ \cite{chakrabarty2019simpler} and more recently to $2+\epsilon$ \cite{ibrahimpur2021minimum}, which remarkably almost matches the bound for makespan. These results introduce  several new ideas to handle general norms by relating them to the special class of top-$k$ norms\footnote{The top-$k$ norm of a non-negative vector $v$ is the sum of its largest $k$ entries.}, and algorithmic techniques to work with them.

\medskip
{\bf General inner and outer norms.}
An even further generalization was considered by Deng, Li and Rabani \cite{deng2023generalized}, which we refer to as the generalized load-balancing (GLB) problem.
Here, additionally, each machine $i$ also has an arbitrary symmetric monotone norm
 $\psi_i:\R^n\rightarrow \R_{\geq 0}$ referred to as the {\em inner-norm},
that determines the machine's load as a function of the sizes of the jobs assigned to
it (note that all the results described previously have inner norm $\ell_1$). 
Formally, given an assignment $\schedule:[n] \rightarrow [m]$ of jobs to machines, the load on machine $i$ is $\Load(i)= \psi_i(p_i^\schedule)$ where $p_i^\schedule$ is the vector of sizes of jobs assigned to $i$, i.e.,~$p_i^\schedule(j)=p_{ij}$ if $\schedule(j)=i$ and $0$ otherwise. The overall objective is $\phi(\Load)$, where $\Load = (\Load(1),\ldots,\Load(m))$ is the vector of machine loads (where $\phi:\R^m\rightarrow \R_{\geq 0}$ is the norm as in \cite{chakrabarty2019approximation}).

 We shall refer to $\phi$ as the {\em outer-norm}.
Throughout this paper a general norm always means a symmetric monotone norm.

In its full generality (when the $\psi_i$ and $\phi$ are general), GLB becomes $\Omega(\log n)$ hard to approximate, as it generalizes\footnote{Given subsets $S_1,\ldots,S_m$ of $[n]$, consider the scheduling instance on $m$ machines (one per set) and $n$ jobs (one per element), with $p_{ij}=1$ iff element $j \in S_i$ and $p_{ij}=\infty$ otherwise. That is, only jobs in $S_i$ can be assigned to $i$. The point is that as $\psi=\ell_\infty$, we have $\Load(i)=0$ if no job is assigned to $i$, and exactly $1$ otherwise (even if all jobs in $S_i$ are assigned to $i$). As $\phi=\ell_1$, the objective is exactly the number of machines (sets) needed to cover all the jobs.}  the Set Cover problem when $\psi=\ell_\infty$ and $\phi=\ell_1$.
Interestingly, \cite{deng2023generalized} gave a matching 
$O(\log n)$ approximation for general $\psi$ and $\phi$, based on
solving and rounding a novel configuration LP.

\medskip

{\bf Generalized Makespan Problem (GMP).}
Given the $\Omega(\log n)$ hardness for the general case, 
\cite{deng2023generalized} also consider the interesting and natural special case where the outer-norm $\ell_\infty$, but the inner norm is general (i.e.,~the goal is to minimize makespan but where the machine loads are given by general inner-norms $\psi_i$). 
We refer to this problem as the {\em Generalized Makepsan Problem} (GMP).
In a sense, this problem can be considered as a ``dual'' of the problem considered by Chakrabarty and Swamy (where the inner-norm is $\ell_1$, but the outer-norm is general).

For GMP, 
Deng, Li and Rabani gave a $3$-approximation for the special case when each $\psi_i$ is a top-$k$ norm. The main open question they ask is whether an $O(1)$ approximation is achievable for general inner-norms. Apriori this seems quite plausible as there is close connection between top-$k$ norms and general norms (see e.g.~\cite{chakrabarty2019approximation,chakrabarty2019simpler}). Moreover, the $O(1)$ approximation of Chakrabarty and Swamy \cite{chakrabarty2019approximation} for the ``dual'' problem, also suggests that GMP may have an $O(1)$ approximation.

\subsection{Our Results}
Our main result is that GMP does not admit an $O(1)$-approximation under standard complexity-theoretic assumptions, answering the main open problem in \cite{deng2023generalized}.

Our starting point is an integrality gap instance for the natural configuration LP for GMP.
\begin{restatable}{theorem}{integralitygap}
\label{thm1}
There is an instance of GMP with a symmetric monotone norm $\psi$, for which the natural configuration LP has an integrality gap of $\Omega((\log n)^{1/2})$.
\end{restatable} 
The gap instance is based on a probabilistic construction and a key idea is to work with a suitably chosen norm $\psi$, defined as the sum of top-$k$ norms at various different scales, that interacts nicely with properties of random set cover instances. This construction is described in \Cref{sec:int-gap}, and it forms a key gadget in our following hardness result.

\begin{restatable}{theorem}{hardnessresult}
\label{thm2}
There is a universal constant $\delta>0$, such that any polynomial-time
 approximation algorithm for GMP has approximation ratio $\Omega(\log^{\delta} n)$ provided that $\emph{\textsf{NP}} \not\subseteq \emph{\textsf{ZTIME}}(n^{O(\log \log n)})$.
\end{restatable}
 
Our construction for this hardness result builds on the ideas of Lund and Yannakakis \cite{lund1994hardness}, who showed the $\Omega(\log n)$
hardness of Set Cover by a gap reduction from the Label Cover problem.
However, we require some additional ideas as reducing the Label Cover problem to a scheduling instance and exploiting the properties of the norm $\psi$ requires much more care. Specifically, even though our gadget in \Cref{thm1} has
the $\Omega((\log n)^{1/2})$ gap, embedding it into Label Cover imposes extra constraints on the number of labels in the Label Cover instance, and only leads to an $\Omega(\log^{\delta} n)$ hardness, for some constant $\delta>0$. 
Interestingly, this constant $\delta$ depends on the soundness parameter of label cover as a function of the number of labels, based on Raz's parallel repetition theorem \cite{raz1995parallel} and its subsequent improvements \cite{holenstein2007parallel,rao2008parallel}. 

\medskip

{\bf Remark.}
The constant $\delta$ can be computed explicitly, but we do not attempt to do this here. This construction is described in \Cref{sec: reduction}. We remark that there are extensive work on improving the soundness in PCP constructions as a function of the number of labels. The best known result in this direction is due to Chan \cite{Chan16} that achieves soundness roughly $L^{-1/2}$ for $L$ labels. However, Chan's result does not have perfect completeness and hence cannot be used in our constructions. Roughly speaking, this is because each job must be assigned to some machine (this is similar to the  reason that one requires perfect completeness in reducing label cover to set cover, as each element must be covered). 

\medskip

{\bf Remark.} In personal communication, Amey Bhangale has pointed out that in an unpublished manuscript, they can show that assuming the $2$-to-$2$ conjecture with perfect completeness, there is a label cover instance with $L$ labels that has perfect completeness and soundness $L^{-1/2}$. Using a such a label cover, our construction in \Cref{sec: reduction} would imply a hardness of $\Omega(\log^{1/8} n)$ in \Cref{thm2}.
We note that currently, proving the $2$-to-$2$ conjecture with perfect completeness remains open, and in particular the breakthrough results of Khot, Minzer and Safra \cite{KhotMS18} on the $2$-to-$2$ conjecture assume imperfect completeness.

%% file: prelims.tex
\section{Preliminaries}

\subsection{\texorpdfstring{$(n, m,\ell, \beta)$}{(n, m, l, beta)} Set-System}
The constructions of the integrality gap instance in \Cref{sec:int-gap} and the reduction from label cover to GMP presented in \Cref{sec: reduction} both use the following set system as a building block. 
\begin{definition}[$(n, \m,  \ell, \bt)$ Set-system] \label{def: mlb-set-system}Let $n, m,\ell$ be positive integers, $\beta \in (0,1)$, $U$ be a set with $|U| = n$, and $A_1, \ldots,  A_{\m}$ be subsets of $U$. The sets $(U; A_1, \ldots, A_m)$ form an $(n, \m,  \ell, \bt)$ set-system if for every set $I$ of at most $ \ell$ indices from $[\m]$, $\left\lvert\cup_{i \in I} B_i\right\rvert \leq (1 - \bt) |U|$, where $B_i$ is either $A_i$ or $\overline{A}_i$. 
\end{definition}

Intuitively, an $(n, m, \ell, \beta)$ set-system has the property that any set cover which uses at most $\ell$ subsets necessarily uses a complementary pair of subsets $A_i$ and $\overline{A}_i$. Moreover, any collection of at most $\ell$ subsets that do not contain any complementary pair can cover at most a $(1-\beta)$ fraction of the elements in $U$. The following lemma shows that for a particular choice of parameters $n, m,\ell, $ and $\beta$, there is a simple and efficient randomized construction of an $(n, m, \ell, \beta)$ set-system.

\begin{lemma}\label{lem: mlb-set-system}
    For a sufficiently large positive integer $n$ and a positive integer $m \in [\sqrt{\log n}, 2\sqrt{\log n}]$ there exists an $(n, \m, \ell,\bt)$ set-system $(U; A_1, \ldots, A_m)$
 with $\,\,\ell = m/10$ and $\bt = \exp(-m) = \exp(-O(\sqrt{\log n}))$. There is a polynomial-time algorithm that constructs such a set system with high probability.
\end{lemma}
\begin{proof}
    Let $U$ be a set of $n$ elements, and initialize $m$ empty sets $A_1, \ldots, A_m$. For each element $e \in U$, sample a random index set $J \subset [\m]$ of size exactly $m/2$ and add $e$ to sets $A_j$ for $j \in J$.

We show that this construction gives an $(n, m,  \ell, \bt)$ set-system with high probability.  
    Consider an index set $I \subset[m]$ with $|I| = \ell$ and a collection of sets $B_i$ for $i \in I$ such that each $B_i$ is either $A_i$ or $\overline{A}_i$. For a fixed $e \in U$, let $p$ denote the probability that $e$ is not contained in $\cup_{i \in I} B_i$, i.e., $p = \Pr[e \notin \cup_{i \in I} B_i]$. We have,
    \[
         p \geq \binom{m-\ell}{m/2}/\binom{m}{m/2}
        \geq \left(\frac{m-\ell}{m/2}\right)^{m/2}/\left(\frac{em}{m/2}\right)^{m/2} 
        \geq \left(\frac{m-\ell}{em} \right)^{m/2}
        \geq \exp(-0.6 m),
    \]
where the second inequality uses that  $\left(\frac{n}{k}\right)^k \leq \binom{n}{k} \leq \left(\frac{en}{k}\right)^k$.

The probability that  $\cup_{i \in I}B_i$  contains a fixed subset of cardinality greater than or equal to $(1-\beta)n$ is at most $(1 - p)^{(1-\beta)n}$. By a union bound over at most $ \binom{n}{\beta n}n$ possible subsets with cardinality at least $(1-\beta)n$,
    \begin{align*}\Pr\left[\lvert \cup_{i \in I}B_i \rvert \geq (1 - \bt) n\right] &\leq \binom{n}{\bt n} n \cdot (1 - p)^{(1-\bt)n}
    \leq \left(e/\bt\right)^{\bt n}n \cdot e^{-(1-\bt)np}\\
    &\leq \exp\left(3n\bt  \log(1/\bt) - np/2\right) \leq \exp(-np/4) \leq \exp(-n^{0.9}),
    \end{align*}
where in the second to last inequality we use that $3\bt \log(1/\bt) < p/4$.

A union bound over the at most $2^\ell \cdot \binom{m}{\ell} \leq \exp( \sqrt{\log n})$ possible choices to pick the $\ell$ sets  $B_i$, gives that with high probability, the union of any $\ell$ sets $B_i$ has cardinality less than $(1-\bt)n$. 
\end{proof}

\subsection{Label Cover}
In \Cref{sec: reduction}, we prove the hardness of approximation of GMP via a reduction from the standard label cover problem as defined below. 
\begin{definition} \label{def: label-cover}
A \emph{label cover} instance $\mathcal{L}$ is defined by a tuple $((U,V, E), \labels, \Pi)$. Here $(U,V,E)$ is a bipartite graph with vertices $U \cup V$ and edges $E \subseteq U \times V$; $\labels$ is a positive integer and $\Pi$ is a set of functions one for each edge $e \in E$ i.e., $\Pi = \{\pi_e: [\labels] \rightarrow [\labels] \,\, |\,\, e \in E \}$. A labeling of the vertices $\sigma: U \cup V \rightarrow [\labels]$ is said to satisfy an edge $e = (u,v)$ if $\pi_e(\sigma(u)) = \sigma(v)$. Given $\mathcal{L}$, the goal of the label cover problem is to find a labeling $\sigma^*$ that satisfies the maximum number of edges in $E$. We use $OPT(\mathcal{L})$ to denote the fraction of the edges in $E$ satisfied by $\sigma^*$.
\end{definition}

As we will need the explicit dependence between the number of labels and the soundness, for completeness we 
sketch below the precise gap version of the label cover problem that we will use. 
\begin{lemma}[Hardness of  Gap  Label Cover] \label{lem: hardness-of-label-cover}
Given a label cover instance $\mathcal{L} = ((U,V, E), \labels, \Pi)$ satisfying:
\begin{enumerate}
  \item[(i)] $|U| = |V| = \lsize$
  \item[(ii)]  The degree of every vertex in $U \cup V$ is $d = O((\log \lsize)^{c_1})$ for some constant $c_1$.
  \item[(iii)]
  $\labels = \sqrt{\log \lsize}$
\end{enumerate}
There is some constant $\razconstant > 0$, for which there is no polynomial-time algorithm to decide if $OPT(\mathcal{L}) = 1$ or $OPT(\mathcal{L}) \leq (\log \lsize)^{-\razconstant}$ provided that $\emph{\textsf{NP}} \not\subseteq \emph{\textsf{DTIME}}(\lsize^{O(\log \log \lsize)})$.
\end{lemma}
\begin{proof}
Using a standard argument (see for ex., \cite{feige1996threshold, arora1996hardness}) one can obtain a reduction from a 3SAT-5 instance $\phi$ with $\satvars$ variables to a Label Cover instance $\mathcal{L}_1 = ((U_1, V_1, E_1), 8, \Pi)$, where $|U_1| = |V_1|=O(\satvars)$ and the graph $(U_1, V_1, E)$ is $15$-regular. The  instance $\mathcal{L}_1$ has the following property: if $\phi$ has a satisfying assignment then $OPT(\mathcal L_1) = 1$; else if any assignment satisfies at most $(1-\epsilon)$ fraction of clauses in $\phi$, then $OPT(\mathcal{L}_1) \leq (1 - \Theta(\epsilon))$. By the PCP-theorem \cite{10.1145/278298.278306} it follows that, for some constant $\epsilon_0 > 0$, deciding if $OPT(\mathcal L_1) = 1$ or $OPT(\mathcal L_1) \leq 1 - \epsilon_0$ is \textsf{NP}-hard.

The following well-known construction \cite{arora1996hardness}
gives stronger inapproximability results for label cover. We define the $k$th power of the label cover instance $\mathcal{L}_k = ((U_k,V_k,E_k), 8^k, \Pi^k)$, where $U_k$, $V_k$ are $k$-tuples of vertices in $U_1$, $V_1$ respectively, $E_k$ is the set of all $k$-tuples of edges in $E_1$. The resulting graph has $N = \satvars^{O(k)}$ vertices and is $(15)^k$-regular. The new set of labels\footnote{The labels are essentially numbers from $1$ to $8^k$.} consist of $k$-tuples of $\{1,\ldots,8\}$.  For an edge $e = (e_1, \ldots, e_k) \in E_k$, we define the function $\pi_e^k(a_1, \ldots, a_k) = (\pi_{e_1}(a_1), \ldots, \pi_{e_k}(a_k))$. Raz's Parallel Repetition Theorem \cite{raz1995parallel}, shows that for the label cover instance constructed above, there exists a constant $\alpha$ such that $OPT(\mathcal L_k) \leq (OPT(\mathcal L_1))^{\alpha k}$. 

We now pick $k$ so that $L = \sqrt{\log N}$. Since $L = 8^k$ and $N = \satvars^{O(k)}$, this gives $k = \Theta(\log \log \satvars)$. This choice of $k$ ensures that $d = (15)^k =(\log N)^{c_1}$ for some constant $c_1$. Moreover, if $OPT(\mathcal{L}_1) \leq (1-\epsilon_0)$, then $OPT(\mathcal L_k) \leq (1-\epsilon_0)^{\alpha k} \leq (\log t)^{-c'} \leq (\log N)^{-\razconstant}$ for some positive constants $\razconstant, c'$.
\end{proof}

%% file: integrality-gap.tex
\section{Integrality Gap for Configuration LP}\label{sec:int-gap}

We begin by describing the configuration LP for GMP. We then
explain the high-level ideas behind the gap construction and the properties we need from the norm $\psi$. We then describe the norm $\psi$ and  the integrality gap instance formally and then prove \Cref{thm1}.
As mentioned earlier, this gap instance and the norm $\psi$ form the key gadget in our hardness construction in \Cref{subsec: reduction}, and understanding it is crucial to the results in \Cref{sec: reduction}.

{\bf Configuration LP.} The most natural LP relaxation of GMP is to consider assignment variables $x_{ij} \in [0,1]$ that determine the fraction of job $j$ assigned to machine $i$, and impose natural constraints. However, simple examples show that such an LP is too weak to handle general norms $\psi$. A stronger relaxation
is the configuration LP, where we have exponentially many variables $x_{i,C}$, one for each machine $i$, and a possible subset of jobs $C$ that can be feasibly assigned to machine $i$.  

Let $J$ denote the set of jobs, and $M=[m]$ be the set of machines. 
For a machine $i\in M$, and subset $C\subseteq J$, let $p_i[C] = (p_{ij}\cdot \mathbbm{1}[j\in C])_{j\in J}$ denote the vector of sizes of jobs in $C$.
Let $T$ be a guess on the optimum makespan (we can do a binary search on $T$).
Call a configuration $C$ {\em valid} for machine $i$ if the load $\psi_i(p_i[C])$ of $C$ on $i$ is at most $T$.
 We will slightly abuse notation and denote $\psi_i(p_i[C])$ by $\psi_i(C)$. 

Consider the following feasibility LP.
\begin{subequations}\label{ConfigLP}
\begin{align}
    \sum_{C\subseteq J} x_{i, C} &\leq 1 && \forall i\in M \label{LP:machine}\\
    \sum\limits_{i\in M, C: j\in C} x_{i,C} &= 1 && \forall j\in J \label{LP:job}\\
     x_{i,C} &= 0 &&\text{if } \psi_i(C) > T \label{LP:threshold}
\end{align}
\end{subequations}
The first constraint says that each machine has at most one configuration. The second constraint says that each job is assigned to a machine, and the last constraint ensures that only valid configurations are considered and hence the generalized makespan on any machine is at most $T$.
This LP can be solved efficiently for any desired accuracy $\epsilon>0$ (so the configurations satisfy $\psi_i(C) \leq (1+\epsilon) T$), see e.g.,~\cite{deng2023generalized}. 

\medskip

{\bf The main idea.}
We start with a simple instructive example,
that motivates our choice of the norm and the choice of parameters that lead to the $\Omega(\sqrt{\log n})$ integrality gap.
Consider a random set system on a universe $U$ of $n$ elements and $m$ sets $A_1,\ldots,A_m$ where each element independently lies in $m/2$ randomly chosen sets. 
We will set $m\ll \log n$.
Create an unrelated machine (in fact restricted assignment) scheduling instance $I$ with $m$ machines, where only the jobs in $A_i$ can be assigned to machine $i$ (all other jobs have infinite size on $i$). As each element lies in $m/2$ sets, the LP solution that picks the configuration $A_i$ on each machine $i$ with $x_{i,A_i}=2/m$ is feasible, and uses only $2$ machines in total.

However, in any integral solution, we claim that several machines must pick a non-trivial fraction of jobs from their sets $A_i$. Roughly, this is because the union of any $\ell\ll m$ sets $A_i$ still leaves about $2^{-\ell} |U|$  uncovered (as each element lies in a set with probability $1/2$). Hence in any feasible integral solution, for any $\ell>0$, at least $\ell$ machines must be assigned at least $2^{-\ell} |U|/m$ jobs.

To exploit this, suppose we define 
$\psi$ as the top-$k$ norm with $k\approx \Omega(2^{-\ell}n/m)$, so that any machine with $\geq k$ jobs incurs the same load, say $T$. Then in any integral solution, at least $\ell$ machines have load $T$, while fractionally at most $2$ machines (in total) have load $T$.
By creating $ m/2$ disjoint copies $I^{(1)},\ldots, I^{(m/2)}$ of these set-cover instances, one would then expect that fractionally each machine has load $T$, while integrally the average load becomes $\Omega(\ell)$.

Unfortunately, this does not quite work as stated above, because once there are several instances $I^{(1)},\ldots, I^{(m/2)}$, as these jobs share the same machines, an algorithm can find a low makespan even if it cannot figure out the good underlying set cover solution.
In fact, this is provably so, as \cite{deng2023generalized}
gave a $3$-approximation when $\psi$ is a $\Top{k}$ norm. 

However, our key observation is that this idea can still be made to work by choosing the instances $I^{(1)},\ldots,I^{(m/2)}$ at different scales (of the number of jobs and processing times) and defining the norm $\psi$ as a suitable mixture of the top-$k$ norms at these different scales.
Roughly, this norm $\psi$ still behaves as a top-$k$ norm at each individual scale, but when jobs from different scales are combined, it takes on a large value, which be used  to create a large gap in the reduction above.

Implementing this idea requires separating every two scales by $\Omega(2^{\ell})$. So the $\ell$ scales leads to instances with size about $2^{\ell^2}$ leading to the choice of $\ell = \Theta(\sqrt{\log n})$ to produce the $\Omega(\ell)$ gap. 
We now give the details.

\medskip

\textbf{The Instance.} The instance will have $n$ jobs and $m = \sqrt{\log n}$ machines. 
We first create $h = m/8$  disjoint set-cover instances $I^{(1)},\ldots, I^{(h)}$, where
 $I^{(s)} = (U(s); A_1(s), A_2(s), \ldots, A_m(s))$ forms an $(n_s, m, \ell, \beta)$ set-system with $\ell = m/10, \beta = \exp(-m)$ and $n_s = (\sqrt{n}/2) \cdot \exp(4ms) = (\sqrt{n}/2)\bt^{-4s}$.
 Note that $n_s$ increases as $\beta^{-4s}$ with $s$, and $n_1 \geq \sqrt{n}$ and $n_h =n/2$, and it is easily checked that parameters for each $I^{(s)}$ satisfy the condition in  \Cref{lem: mlb-set-system}.

For each $s\in[h]$, the elements in $U(s)$ correspond to jobs with size
$p^{(s)}=\bt^{s-1}$  (or infinite if the job cannot be assigned to a machine). Each job $j$ in $U(s)$ has size $p_{ij} = p^{(s)}$ on machine $i$ iff $j \in A_i(s)$ and $p_{ij}=\infty$ otherwise.
We abuse the notation slightly to refer to the resulting scheduling instance also as $I^{(s)}$.
As the instances $I^{(1)}, \ldots, I^{(h)}$ are at different scales in terms of the number of jobs and processing times, we refer to jobs in $I^{(s)}$ as being in the $s$-th \textit{size class} 
\footnote{The total number of jobs in the $h$ instances is $\sum_{i = 1}^h n_s \approx n/2$. To make the total number of jobs $n$,  we add dummy jobs that have processing time $0$ on all machines.}.

We define the inner norm $\psi = \sum_{s\in [h]}\term{s}$, where each $\term{s}$ is a scaled top-$k$ norm given by
\begin{equation}\label{innernorm}
    \term{s}(\normvar) = \frac{\Top{\bt^2n_s}(\normvar)}{(\bt^{2}n_sp^{(s)})}.
\end{equation}

Each machine $i$ will have the same inner norm $\psi_i=\psi$.
 Informally, $\psi$ has the following key property: for any subset $C$ of $A_i(s)$ with at least $\beta^2 n_s$ jobs, $\term{s}_i(C) = 1$ and $\term{s'}_i(C) \approx 0$ for $s' \neq s$. This in particular implies that $\psi_i(A_i(s)) \approx 1$ for any class $s$. Moreover, if $C$ is an arbitrary subset of jobs with at least a $\beta^2$ fraction of jobs from $r$ distinct sets among $A_i(1), A_i(2), \cdots, A_i(h)$, then $\psi_i(C) \approx r$  (roughly, each such size-class $s$ affects a different term $\psi^{(s)}$ of the norm). This property motivates the following definition.

\begin{definition}[Heavy Size Class]\label{def:heavysets}
Given an assignment of jobs $\schedule: {J}\rightarrow {M}$, we say that size class $s\in[h]$ is heavy on machine $i\in M$, if at least $\bt^2n_s$ jobs from $A_i(s)$ are assigned to machine $i$. 
\end{definition}
We now formally state and prove
the property of the norm described above.

\begin{lemma}\label{lem:heavy_sets_bound}
   For any  $s\in [h]$ and $i\in M$, $\psi_i(A_i(s)) = 1+o(1)$. Furthermore, for an assignment $\schedule:J\rightarrow M$, if $C$ is the set of jobs assigned to machine $i$, then  $\psi_i(C)$ is at least the number of heavy size classes $s\in [h]$ on machine $i$.
\end{lemma}
\begin{proof}
We first show that $\psi_i(A_i(s)) = 1+o(1)$ by computing the value of $\term{s'}$ for different $s'$. By the definition of $\term{s'}$,
    \begin{align*}
    \term{s'}_i(A_i(s)) = \frac{\Top{\bt^2n_{s'}}(p_i[A_i(s)])}{(\bt^2n_{s'}p^{(s')})}  = \frac{\min\{\bt^2n_{s'},|A_i(s)|\}\cdot p^{(s)}}{(\bt^2n_{s'}p^{(s')})} = \left(\min\left\{1,\frac{|A_i(s)|}{\bt^2n_{s'}}\right\}\right)\frac{p^{(s)}}{p^{(s')}}
    \end{align*}
    For $s=s'$, this exactly equals $1$ (as $\beta\ll 1$ and $|A_i(s)|\approx \frac{n_s}{2}$).
    
\noindent For $s'<s$, we have   $\term{s'}_i(A_i(s)) \leq \frac{p^{(s)}}{p^{(s')}} = \bt^{(s-s')} \leq \bt.$

\noindent Finally, for $s'>s$,  we have
\[ \term{s'}_i(A_i(s))  \leq  \frac{|A_i(s)|}{\bt^2n_{s'}}\cdot\frac{p^{(s)}}{p^{(s')}}  \leq  \frac{n_s}{\bt^2n_{s'}}\cdot\frac{p^{(s)}}{p^{(s')}} = \bt^{3(s'-s)-2} \leq \bt.\]

Let us now consider an arbitrary set of jobs $C$ assigned to machine $i$. By the monotonicity of the norm,
    \[     \term{s}_i(C) \geq \term{s}_i(C\cap A_i(s)) = \min\left\{1,\frac{|C\cap A_i(s)|}{\bt^2n_{s}}\right\} \geq \mathbbm{1}[s \text{ is heavy on }i].    \]
    As $\psi_i(C) = \sum\limits_{s\in [h]}\term{s}_i(C)$, 
it follows that $\psi_i(C)$ is at least the number of heavy size classes on $i$.
\end{proof}

\medskip

\noindent{\bf Fractional Solution.} 
 
We claim that the following solution is feasible for the configuration LP \Cref{ConfigLP} with $T=2$:
 for each machine $i\in M$, set $x_{i, A_i(s)} = 2/m$  for each $s\in [h]$.
 
Clearly, \Cref{LP:machine} is satisfied for each $i\in {M}$ as there are $h<m/2$ sets of jobs $A_i(1), A_i(2), \ldots, A_i(h)$, each with value $2/m$. \Cref{LP:job} is satisfied as each job $j\in U(s)$, for each $s\in [h]$, lies in $m/2$ sets in $A_1(s), A_2(s), \ldots, A_m(s)$. \Cref{LP:threshold} is also satisfied as $\psi_i(A_i(s)) = 1+o(1) <2=T$ by \Cref{lem:heavy_sets_bound}.

\medskip

\noindent{\bf Integral Solution.}
We now show that any integral solution has generalized makespan  $\Omega(\sqrt{\log n})$.

\begin{lemma}
For any assignment $\schedule: J \rightarrow M$, there is some machine with load $\Omega(\sqrt{\log n})$. 
\end{lemma}
\begin{proof}
We first show that each size class is heavy on at least $\ell=\sqrt{\log n}/10$ machines. 
Suppose this is not true for some size class $s\in [h]$. Let $H\subseteq M$ be the set of machines on which $s$ is heavy, and so $|H|<\ell$. As $I^{(s)}$ forms an $(n_s, m, \ell, \bt)$ set-system, by \Cref{def: mlb-set-system}, $|\cup_{i\in H} A_i(s)| \leq (1-\beta)n_s$, and hence at least $\bt n_s$ jobs from $U(s)$ are assigned to machines $i\notin H$. However, as any machine $i\notin H$ can have at most $\bt^2 n_s$ jobs from $A_i(s)$, the machines in $i\notin H$ can have is at most $m\cdot \bt^2n_s < \bt n_s$, contradicting that each job in $U(s)$ was assigned to some machine.

By averaging over machines, there exists a machine $i$ on which at least $(h\ell/m) = \Omega(\sqrt{\log n})$ size classes are heavy. By \Cref{lem:heavy_sets_bound}, this implies that $\psi_i = \Omega(\sqrt{\log n})$.
\end{proof}
This concludes the proof of \Cref{thm1}.

%% file: reduction.tex
\section{Reduction from Label-Cover}\label{sec: reduction}
We now prove  \Cref{thm2} by reducing Label Cover to a GMP instance. 

\medskip

{\bf Overview.}
Our construction builds on the ideas used by Lund and Yannakakis \cite{lund1994hardness} to show the $\Omega(\log n)$ hardness of set cover, using a gadget based on a natural $\Omega(\log n)$ integrality gap instance. We first give a rough sketch of their idea (see e.g.~\cite{arora1996hardness} for an excellent exposition), and then explain the additional steps needed in our setting and why we only get a $\log^\delta n$ hardness for some small $\delta>0$, despite the $\Omega(\log^{1/2}n)$ integrality gap instance above.

Consider a label cover instance $\mathcal{L} = ((U,V,E),L,\Pi)$, as defined in \Cref{def: label-cover}, with label set $[L]$ and vertex sets $U$ and $V$. The main idea in \cite{lund1994hardness} is the following. For each edge $e =(u,v)\in E$, one associates a $(n,m,\ell,\beta)$-system $I^e =(U^e; A_1^e \ldots, A_m^e)$  (with disjoint universes for each edge). The sets will be associated with labels for vertices (so that $L=m$) and we associate the set $A_{\pi_e(a)}^e$ with label $a$ for $u$ and $\overline{A}_b^e$ 
 with label $b$ for vertex $v$. The point is that in the completeness case, where the labels $a'$ and $b'$ for $u$ and $v$ satisfy $e$ (i.e.,~$\pi_e(a')=b'$), $U^{e}$ can be covered by just the two corresponding sets 
$A_{\pi_e(a')}^e$ and $\overline{A}_{b'}^e$. 
Conversely, in the soundness case, if $U^e$ is covered using less than $\ell$ sets, there must be a pair of sets that are complements of each other, which can be used to produce a good labeling for $\mathcal{L}$.

To adapt this to our setting, suppose we create a job for each element in $U^e$, and $m$ machines per vertex, one for each label in $[L]$. Also suppose that for labels $a,b$, we set the processing times of jobs in the set $A_{\pi_e(a)}^e$ to be finite on the $a$-th machine of vertex $u$ and the processing times of jobs in set $\overline{A_{b}^e}$ to be finite on the $b$-th machine of vertex $v$.
In the completeness case, any perfect labeling gives an assignment where jobs are assigned to exactly one out of the $m$ machines per vertex (similar to the value of the LP solution in the gap example in \Cref{sec:int-gap}).
However, the soundness argument fails, as a low makespan assignment can spread jobs from $U^e$ on multiple machines, and not give any information to recover a good labeling 
(this is similar to the reason we needed multiple size classes in the gap example in \Cref{sec:int-gap}).

To get around this, for each $e$, we will use 
$h$ different set systems $I^{e,1},\ldots,I^{e,h}$ (of geometrically increasing sizes) where the $s$-th set system is the following $I^{e,s} = (U^e(s); A_1^e(s), \ldots, A_m^e(s))$.
Each vertex will have $m$ machines, one for each label. The intended solution is that if vertex $u$ is assigned label $a$, then we pick the sets $A^e_{\pi_e(a)}(1),\ldots,A^e_{\pi_e(a)}(h)$, and assign the corresponding jobs to   the $m$ machines for $u$ with small makespan (this requires some care so that for any label $a$, jobs from different classes $s$ can be assigned to different machines, but we ignore this issue here).

Using the properties of the norm $\psi$ and the arguments in \Cref{sec:int-gap} for the integrality gap, one can show that if $h=\Omega(m)$, then given any schedule with low makespan,
one can construct a good labeling thereby proving soundness. A key new idea here beyond \cite{lund1994hardness} is to show that there is some fixed size class $s^*$, such that the assignment of jobs in class $s^*$ gives a small set of good candidate labels for a large fraction of edges.
However, as the hardness of Label-Cover is only $\Omega(L^c)$ for some small constant $c$ as a function of the number of labels, our resulting hardness is only $\Omega(\log^{c'}n)$ for some small $c'>0$, instead of $\Omega(\log^{1/2} n$).

\subsection{The Reduction}\label{subsec: reduction}
Suppose that we are given a label cover instance, $\mathcal{L} = ((U, V, E), \labels, \Pi)$ satisfying the properties of \Cref{lem: hardness-of-label-cover}, i.e., the number of vertices $|U| = |V| = \lsize$, the degree of every vertex is $d = O((\log \lsize)^{c_1})$ and the number of labels $L = \sqrt{\log \lsize}$. We now describe a polynomial-time (randomized) reduction from $\mathcal{L}$ to a GMP instance $\instance$ with machines ${M}$, jobs ${J}$ and assign processing times for each machine $i\in{M}$ and job $j\in {J}$. 

\medskip

\noindent\textbf{Machines.} For each vertex $w\in U\cup V$, we create $m=\labels = \sqrt{\log \lsize}$ machines. We denote the set of machines corresponding to vertex $w$ by $M_w = \{\mach{1}{w}, \ldots, \mach{m}{w}\}$. We denote the set of all machines by ${M}=\bigcup_{w\in U\cup V}M_w$. In total, we have $2N \sqrt{\log N}$ machines.

\medskip

\noindent\textbf{Jobs.} For each edge $e \in E$, we create $O(N)$ jobs and partition them into $h=m/8$ size-classes  $U^e(1), U^e(2), \ldots, U^e(h)$ of geometrically increasing size. More precisely, we pick the number of jobs in the $s$-th set to be $|U^e(s)| = \sqrt{N} \cdot\exp(4s\sqrt{\log N})$ which is always $O(N)$ since $s \leq \sqrt{\log \lsize}/8$. Therefore, the total number $n$ of jobs created is $\poly(N)$. We also remark that these sets can be constructed efficiently by the randomized procedure described in \Cref{lem: mlb-set-system} and that this is the only randomized step of the reduction.

\medskip

\noindent\textbf{Processing times.} To assign processing times, for each edge $e\in E$ and  $s \in [h]$,  we construct a $(|U^e(s)|, m,\ell,\beta)$ set system $I^{e,s} = (U^e(s); A_1^e(s), \ldots, A_m^e(s))$ with $\ell = m/10$ and $\beta = \exp(-m)$. For every vertex $u\in U$, label $a\in [\labels]$ and $s\in [h]$, define the set
\begin{align}\label{eq: suas}
\Set{u}{a}{s} = \bigcup\limits_{e\in \delta(u)} A^e_{\pi_e(a)}(s)\text{.}
\end{align}
Similarly for every vertex $v\in V$, label $b\in [\labels]$ and $s\in [h]$, define the set
\begin{align}\label{eq: svas}
\Set{v}{b}{s} = \bigcup\limits_{e\in \delta(v)} \overline{A^e_{b}}(s)\text{.}
\end{align}
For every vertex $w\in U\cup V$, label $a\in [L]$ and size class $s\in [h]$, we choose the processing times of all the jobs in $\Set{w}{a}{s}$ to be ${p^{(s)} = \bt^{s-1}}$ on the machine $\mach{i}{w}$ if the index $i$ satisfies $i\equiv(a+s)\pmod m$. This way of assigning processing times ensures that for a fixed machine $\mach{i}{w}$ and a label $a$, there is at most one set of jobs among $S_{w,a,1}, S_{w,a,2}, \ldots, S_{w,a,h}$ that have finite processing time on $\mach{i}{w}$. This is a useful property to have when we prove the completeness of our reduction.

\medskip
{\noindent \bf Norm.} Define the norm $\psi = \sum_{s\in[h]} \term{s}$, where each $\term{s}$ is a scaled top-$k$ norm given by
\begin{equation}\label{eq:innernorm-final}
    \term{s}(\normvar) = \frac{\Top{\bt^2 n_s}(\normvar)}{(\bt^2 n_s p^{(s)})}
\end{equation}
where $n_s = \left\lvert\bigcup_{e\in\delta(w)}U^e(s)\right\rvert$ is the number of jobs of size class $s$ contained in edges incident to any vertex. Note that since the graph is $d$-regular this number is the same for each vertex.

The above reduction takes $\poly(N)$ time because there are only polynomially many jobs, machines and the $(n,m,\ell, \beta)$ set systems required can be constructed efficiently by \Cref{lem: mlb-set-system}.

\subsection{Analysis}
The set of jobs $\Set{w}{a}{s}$ for any vertex $w$, label $a$ and size class $s$, only contribute to the $s$-th top-k term of $\psi$ on machine $\mach{i}{w}$. Similar to \Cref{def:heavysets}, we define heavy and light size classes and state a lemma whose proof is analogous to that of \Cref{lem:heavy_sets_bound}.
\begin{definition}[Heavy Size Class]
   For a vertex $w$, label $a$ and size class $s$, consider the set $\Set{w}{a}{s}$ of jobs and the machine $\mach{i}{w}$ satisfying $i \equiv (a + s) \pmod m$. For an assignment $\schedule: {J} \rightarrow {M}$, we say that size class $s$ is heavy on machine $\mach{i}{w}$, if $\schedule$ assigns at least $\beta^2 n_s$ jobs from $\Set{w}{a}{s}$ to machine $\mach{i}{w}$; otherwise, we say the size class $s$ is light on machine $\mach{i}{w}$.
\end{definition}

\begin{lemma}\label{lem: norm-is-nice}
    For any vertex $w$, label $a$, size class $s$,  and a machine $\mach{i}{w}$ satisfying $i\equiv a+s\pmod m$, the norm $\psi_{\mach{i}{w}}(\Set{w}{a}{s}) = 1+o(1)$. Furthermore, for an assignment $\schedule:J\rightarrow M$, let $S$ be the set of jobs assigned to machine $\mach{i}{w}$. Then $\psi_{\mach{i}{w}}(S)$ is at least the number of heavy size classes heavy on $\mach{i}{w}$. 
\end{lemma}

\subsubsection{Completeness}
Given a labeling $\sigma$ for the label cover instance $\mathcal{L}$ which satisfies all the edges, we use it to construct an assignment of jobs $\schedule$ with a low makespan. 
\begin{lemma}\label{lem: completeness}
   If the label cover instance $\mathcal{L}$ satisfies $OPT(\mathcal{L}) = 1$, the  instance $\instance$ has an assignment $\schedule: {J} \rightarrow {M}$ with makespan less than $2$.
\end{lemma}
\begin{proof}
     Let $\sigma$ be the labeling of vertices that satisfies all edges in the label cover instance $\mathcal{L}$. Consider the assignment $\schedule$ of jobs to machines constructed using $\sigma$ in the following way: for every vertex $w\in U\cup V$, and size class $s\in[h]$, assign the jobs in  $\Set{w}{\sigma(w)}{s}$ to machine $\mach{i}{w}$ where the index $i$ satisfies $i\equiv\sigma(w)+s\pmod m$.

    For an edge $e=(u,v)\in E$ we first show that each job in the sets  $U^e(1), U^e(2),\ldots, U^e(h)$ is assigned to some machine. For a size class $s\in[h]$, the jobs in  $\Set{u}{\sigma(u)}{s}$ are assigned to a machine in $M_u$, and similarly, the jobs in $\Set{v}{\sigma(v)}{s}$ are assigned to a machine in $M_v$. Since $\pi_e(\sigma(u)) = \sigma(v)$, we infer that $A^e_{\sigma(v)}(s) = A^e_{\pi_e(\sigma(u))}(s)  \subseteq \Set{u}{\sigma(u)}{s}$ and $\overline{A^e_{\sigma(v)}}(s) \subseteq \Set{v}{\sigma(v)}{s}$. It follows that each job in $U^e(s)$ is assigned to some machine.

    We now bound the makespan of the assignment by showing that each machine is assigned jobs from at most one size class. For a vertex $w$ and $i \in [m]$ consider the machine $w_i$. Since $h < m$, there is at most one $s \in [h]$ for which $\sigma(w) + s \equiv i \pmod m$. Therefore, $\mach{i}{w}$ is assigned jobs from at most one of the sets $\Set{w}{\sigma(w)}{1}, \Set{w}{\sigma(w)}{2}, \cdots, \Set{w}{\sigma(w)}{h}$. Therefore, by \Cref{lem: norm-is-nice} its norm is at most $1+o(1)<2$.
\end{proof}
    
\subsubsection{Soundness}
Next, we show that if the instance $\instance$ has a makespan much less than $\ell$, then $OPT(\mathcal{L})$ is large. Towards this end, we first prove some useful lemmas. Consider an assignment $\schedule$ with makespan $T \leq \ell/100$. Call a size class $s$ to be \textit{good} for a vertex $w$ if it is heavy on at most $32T$ of the machines $\mach{1}{w}, \ldots, \mach{m}{w}$; if not define it to be bad. We first show that a large fraction of size classes are good for any vertex.
\begin{lemma}\label{lem: good-size-classes}
    There are at least $3h/4$ good size classes for each vertex.
\end{lemma}
\begin{proof}
    Let $b$ be the number of bad size classes for some vertex $w$.  
     By averaging over the $m$ machines on vertex $w$, there is a machine $\mach{i}{w}$ for which at least $(32Tb)/m$ size classes are heavy. By \Cref{lem: norm-is-nice},  $\mach{i}{w}$ has  norm at least $(32Tb)/m$. As any machine has norm at most $T$, we get $b \leq m/32 = h/4$ and hence the claim follows.
\end{proof}

Call a size class {\em good} for an edge $(u,v)$ if it is good for both $u$ and $v$. By \Cref{lem: good-size-classes} each edge $e$ has at least $h/2$ good size classes. 
By averaging over the edges $e \in E$, there must exist a size class $s^* \in [h]$ which is good for at least $|E|/2$ edges.

We will fix the class $s^*$ henceforth, and use it to construct a good label cover solution by assigning a suitable label to each vertex. These labels will only depend on the class $s^*$.

\medskip

\noindent{\bf Constructing a good labeling.}
 For each vertex $w \in U \cup V$, we define $L(w)$ to be the set of all labels $a$ such that size-class $s^*$ is heavy on $\mach{i}{w}$ where $i \equiv (a + s) \pmod m$. If no such label exists, add an arbitrary label to $L(w)$.
 \begin{lemma}\label{lem: existence-of-matching-labels}
     Let $e = (u,v)$ be an edge for which $s^*$ is good. There exists $a \in L(u)$ and $b \in L(v)$ such that $\pi_e(a) = b$.
 \end{lemma}
 \begin{proof}
         Assume there are no such labels $a\in L(u)$ and $b\in L(v)$ for which  $\pi_e(a)=b$. Since $|L(u)|+|L(v)|\leq 64T <\ell$, then the union of all the sets $A_{\pi_e(a)}^e(s^*)$ and $\overline{A_b^e}(s^*)$ such that $a\in L(u)$ and $b\in L(v)$ covers at most $(1-\bt)|U^e(s^*)|$ from \Cref{def: mlb-set-system}. 
         
         From the definition of a light size class, for all labels $a\notin L(u)$ (resp. $b\notin L(v)$), at most $\bt^2n_{s^*} = \bt^2 (|U^e(s^*)|\cdot d)$ jobs from the sets $A_{\pi_e(a)}^e(s^*)$ (resp. $\overline{A_b^e}(s^*)$) are assigned to some machines on vertex $u$ (resp. $v$) . Notice that the degree of the graph $(U,V,E)$ is $d = O((\log N)^{c_1})$. So, the union of all the jobs assigned from the sets $A_{\pi_e(a)}^e(s^*)$ (resp. $\overline{A_b^e}(s^*)$) such that $a\notin L(u)$ (resp. $b\notin L(v)$) has at most $(2m\cdot\bt^2\cdot|U^e(s^*)|\cdot d) < \bt|U^e(s^*)|$ jobs which is a contradiction.
\end{proof}
  For each vertex $w$, set $\sigma(w)$ to be a label selected uniformly at random from $L(w)$. We show that $\sigma$ satisfies a large fraction of edges of $\mathcal{L}$ completing the proof of soundness.

\begin{lemma}
\label{lem: soundness}
   If  the  instance $\instance$ has an assignment $\schedule: {J} \rightarrow {M}$ with makespan $T \leq \ell/100$, then $OPT(\mathcal{L}) \geq 1/(2048T^2)$.
\end{lemma}
\begin{proof}
    Consider an edge $e = (u,v)$ for which size class $s^*$ is good. In this case, we have $|L(u)| \leq 32T$ and $|L(v)| \leq 32T$. Also, by \Cref{lem: existence-of-matching-labels} there exists $a \in L(u)$ and $b \in L(w)$ for which $\pi_e(a) = b$. Therefore, $\pi_e(\sigma(u)) = \sigma(v)$ with probability at least $1/(32T)^2$. We have by the analysis above that the number of edges for which $s^*$ is good is at least $|E|/2$. Therefore, the expected number of edges satisfied by $\sigma$ is at least $|E|/(2(32)^2T^2) = |E|/(2048 T^2)$ and the lemma follows.
\end{proof}

\hardnessresult*

\begin{proof}
    Suppose that we have an algorithm that in polynomial time can decide if the $GMP$ instance constructed has a makespan of at least $T$ or at most $2$. By the reduction above, from \Cref{lem: completeness,lem: soundness}  this algorithm can also distinguish between label cover instances that have value $1$ and those that have value at most $1/2048T^2$. Due to the hardness of label cover (\Cref{lem: hardness-of-label-cover}), this is not possible if $1/2048T^2 \geq 1/(\log N)^{\razconstant}$, i.e., if $T \leq O((\log N)^{\razconstant/2}) = O((\log n)^{\razconstant/2})$ since the number of jobs $n=\poly(N)$. This, in particular, implies that any approximation algorithm for GMP has an approximation ratio of at least $\Omega((\log n)^{\razconstant/2})$ provided that $\textsf{NP} \not\subset \textsf{ZTIME}(n^{O(\log \log n)}).$
\end{proof}

%% file: conclusion.tex
\section{Concluding Remarks}
We conjecture that GMP admits an $O(\sqrt{\log n})$ approximation, based on suitably rounding the configuration LP. However, we are unable to prove any $o(\log n)$ approximation even in the restricted assignment case (note that the integrality gap and hardness instances in this paper only use restricted assignment).
Finding a $o(\log n)$ approximation algorithm for any of these variants would be extremely interesting.

%% file: main.bbl
\newcommand{\etalchar}[1]{$^{#1}$}
\begin{thebibliography}{AAWY98}

\bibitem[AAWY98]{alon1998approximation}
Noga Alon, Yossi Azar, Gerhard~J Woeginger, and Tal Yadid.
\newblock Approximation schemes for scheduling on parallel machines.
\newblock {\em Journal of Scheduling}, 1(1):55--66, 1998.

\bibitem[AE05]{azar2005convex}
Yossi Azar and Amir Epstein.
\newblock Convex programming for scheduling unrelated parallel machines.
\newblock In {\em Symposium on Theory of Computing}, pages 331--337, 2005.

\bibitem[AL97]{arora1996hardness}
Sanjeev Arora and Carsten Lund.
\newblock Hardness of approximations.
\newblock In {\em Approximation algorithms for NP-hard problems}, pages
  399--446. PWS Publishing Company, Boston, 1997.

\bibitem[ALM{\etalchar{+}}98]{10.1145/278298.278306}
Sanjeev Arora, Carsten Lund, Rajeev Motwani, Madhu Sudan, and Mario Szegedy.
\newblock Proof verification and the hardness of approximation problems.
\newblock {\em J. ACM}, 45(3):501–555, may 1998.

\bibitem[Cha16]{Chan16}
Siu~On Chan.
\newblock Approximation resistance from pairwise-independent subgroups.
\newblock {\em J. {ACM}}, 63(3):27:1--27:32, 2016.

\bibitem[CS19a]{chakrabarty2019approximation}
Deeparnab Chakrabarty and Chaitanya Swamy.
\newblock Approximation algorithms for minimum norm and ordered optimization
  problems.
\newblock In {\em Symposium on Theory of Computing}, pages 126--137, 2019.

\bibitem[CS19b]{chakrabarty2019simpler}
Deeparnab Chakrabarty and Chaitanya Swamy.
\newblock Simpler and better algorithms for minimum-norm load balancing.
\newblock In {\em 27th Annual European Symposium on Algorithms}, volume 144 of
  {\em LIPIcs}, pages 27:1--27:12, 2019.

\bibitem[DLR23]{deng2023generalized}
Shichuan Deng, Jian Li, and Yuval Rabani.
\newblock Generalized unrelated machine scheduling problem.
\newblock In {\em Symposium on Discrete Algorithms (SODA)}, pages 2898--2916.
  SIAM, 2023.

\bibitem[Fei96]{feige1996threshold}
Uriel Feige.
\newblock A threshold of $\ln n$ for approximating set cover (preliminary
  version).
\newblock In {\em Symposium on the Theory of Computing}, pages 314--318, 1996.

\bibitem[Hol07]{holenstein2007parallel}
Thomas Holenstein.
\newblock Parallel repetition: simplifications and the no-signaling case.
\newblock In {\em Symposium on Theory of Computing}, pages 411--419, 2007.

\bibitem[IS21]{ibrahimpur2021minimum}
Sharat Ibrahimpur and Chaitanya Swamy.
\newblock Minimum-norm load balancing is (almost) as easy as minimizing
  makespan.
\newblock In {\em 48th International Colloquium on Automata, Languages, and
  Programming}, volume 198 of {\em LIPIcs}, pages 81:1--81:20, 2021.

\bibitem[KMPS09]{kumar2009unified}
V.~S.~Anil Kumar, Madhav~V. Marathe, Srinivasan Parthasarathy, and Aravind
  Srinivasan.
\newblock A unified approach to scheduling on unrelated parallel machines.
\newblock {\em J. {ACM}}, 56(5):28:1--28:31, 2009.

\bibitem[KMS18]{KhotMS18}
Subhash Khot, Dor Minzer, and Muli Safra.
\newblock Pseudorandom sets in grassmann graph have near-perfect expansion.
\newblock In {\em Symposium on Foundations of Computer Science, {FOCS}}, pages
  592--601, 2018.

\bibitem[LST90]{lenstra1990approximation}
Jan~Karel Lenstra, David~B. Shmoys, and {\'{E}}va Tardos.
\newblock Approximation algorithms for scheduling unrelated parallel machines.
\newblock {\em Math. Program.}, 46:259--271, 1990.

\bibitem[LY94]{lund1994hardness}
Carsten Lund and Mihalis Yannakakis.
\newblock On the hardness of approximating minimization problems.
\newblock {\em Journal of the ACM (JACM)}, 41(5):960--981, 1994.

\bibitem[MS18]{makarychev2018optimization}
Konstantin Makarychev and Maxim Sviridenko.
\newblock Solving optimization problems with diseconomies of scale via
  decoupling.
\newblock {\em J. {ACM}}, 65(6):42:1--42:27, 2018.

\bibitem[Rao08]{rao2008parallel}
Anup Rao.
\newblock Parallel repetition in projection games and a concentration bound.
\newblock In {\em Symposium on Theory of Computing}, pages 1--10, 2008.

\bibitem[Raz95]{raz1995parallel}
Ran Raz.
\newblock A parallel repetition theorem.
\newblock In {\em Symposium on Theory of computing}, pages 447--456, 1995.

\end{thebibliography}
